\documentclass[prd,aps,tightline,twocolumn,nofootinbib,preprintnumbers]{revtex4}
\usepackage{graphicx}
\usepackage{amssymb}
\usepackage{amsmath}
\usepackage{color}

\preprint{KEK-TH-1428}

\begin{document}
 
\newcommand{\be}{\begin{equation}}
\newcommand{\ee}{\end{equation}}
\newcommand{\aprime}{\mathbf{a}^{\prime}}
\newcommand{\bprime}{\mathbf{b}^{\prime}}
\newcommand{\kh}{\hat{k}}
\newcommand{\Ip}{\vec{I}_+}
\newcommand{\Imi}{\vec{I}_-}
\newcommand{\bc}{\begin{cases}}
\newcommand{\ec}{\end{cases}}
\newcommand{\cD}{\mathcal{D}}
\newcommand{\xv}{\mathbf{x}}
\newcommand{\qv}{\mathbf{q}}
\newcommand{\pv}{\mathbf{p}}
\newcommand{\trec}{t_{\mathrm{rec}}}
\newcommand{\trei}{t_{\mathrm{rei}}}

\newcommand{\red}{\color{red}}
\newcommand{\cyan}{\color{cyan}}
\newcommand{\blue}{\color{blue}}
\newcommand{\magenta}{\color{magenta}}
\newcommand{\yellow}{\color{yellow}}
\newcommand{\green}{\color{green}}
\newcommand{\rem}[1]{{\bf\blue #1}}

%\begin{flushright}
%KEK-TH-1428
%\end{flushright}

%\vskip 1cm

\title{
On the waterfall behavior in hybrid inflation
}

%\vskip 1cm

\author{Hideo Kodama$^{(a,b)}$, Kazunori Kohri$^{(a,b)}$,
and Kazunori Nakayama$^{(a)}$}

\affiliation{%
$^a$Theory Center, Institute of Particle and Nuclear Studies, KEK, 
        1-1 Oho, Tsukuba, Ibaraki 305-0801, Japan \\
$^b$Department of Particles and Nuclear Physics, The Graduate University for Advanced Studies,
        1-1 Oho, Tsukuba, Ibaraki 305-0801, Japan
}

\date{\today}

\vskip 1.0cm

\begin{abstract}
We revisit the hybrid inflation model focusing on the dynamics of the waterfall field in an analytical way.
It is shown that inflation may last long enough during the waterfall regime for some parameter regions,
confirming the claim of Clesse.
In this case the scalar spectral index becomes red, and can fall into the best fit range of the WMAP observation.
\end{abstract}

 \maketitle

%%%%%%%%%%%%%%%%%%%%%%%%%%%%%%%%%%%%%%%%%%%%%% 
 \section{Introduction}
 %%%%%%%%%%%%%%%%%%%%%%%%%%%%%%%%%%%%%%%%%%%%%%

 Inflation~\cite{Guth:1980zm} has become a standard scenario in the early Universe cosmology.
 It not only solves the horizon problem and flatness problem, but also explains
 the primordial density fluctuation of the Universe through the quantum fluctuation of the inflaton, which
 is a scalar field driving the inflationary expansion.
 
There are many models of inflation so far :
new inflation~\cite{Albrecht:1982wi}, chaotic inflation~\cite{Linde:1983gd}, 
hybrid inflation~\cite{Linde:1991km,Linde:1993cn}, and so on.
But it is still a challenging task to pin down the model observationally.
This is partly because observational information is limited.
What we extract from cosmological observations are the 
spectral index of the power spectrum of the density perturbation, $n_S$,
and tensor-to-scalar ratio, $r$.
Currently, the WMAP satellite combined with measurements of baryon acoustic oscillation and the present Hubble parameter
constrains $n_S$ as $n_S = 0.968\pm 0.012$ with 68\% C.L.,
and only an upper bound on $r$ is given as $r < 0.24$ with 95\% C.L.~\cite{Komatsu:2010fb}.
If future observations improve the accuracy of these parameters and find the evidence of primordial tensor perturbation,
it will help us determine the correct inflation model.
But even at the current level, some inflation models begin to be excluded.
Generally, inflation models which predict blue spectral index, $n_S > 1$, are not favored
from observational point of view.

The hybrid inflation is one of the models that predict blue spectral index, and hence people
often consider that hybrid inflation models are on the verge of exclusion.
In the hybrid inflation model, two scalar fields are introduced : one is the inflaton field $\phi$
and the other is the waterfall field $\psi$.
The mass of the $\psi$ is arranged so that it becomes tachyonic at some critical value of $\phi = \phi_c$,
and inflation suddenly ends at this point.
Recently, Ref.~\cite{Clesse:2010iz} argued that this picture might change for some cases,
since inflation with large amount of e-foldings may still occur during the waterfall regime.
In this case it is $\psi$, not $\phi$, that actually takes a role of the inflaton relevant for observational scales.

In this paper we revisit the hybrid inflation model, focusing on the
detailed dynamics of the waterfall field.  While
Ref.~\cite{Clesse:2010iz} numerically investigated this topic, we
analytically study the model in detail.  We confirmed the claim of
Ref.~\cite{Clesse:2010iz} and reproduced their results.  For some
parameter spaces, the waterfall field $\psi$ causes the final 60
e-foldings of the inflationary expansion, which resembles new
inflation or type I hilltop inflation scenario~\cite{Kohri:2007gq,Boubekeur:2005zm},
and in this case the spectral index becomes red, on the contrary to
the lore that the hybrid inflation predicts the blue spectral index.
 
The effects of the fluctuation of the waterfall field are also discussed recently~\cite{Lyth:2010ch,Abolhasani:2010kr,Fonseca:2010nk,Gong:2010zf,Abolhasani:2010kn,Lyth:2010zq}.
In these studies, most of the 60 e-foldings are assumed to be spent before the waterfall regime.
On the other hand, we are interested in the case that 60 e-foldings occurs during the waterfall regime.
Thus our analysis is based on the classical dynamics of the $\phi$ and $\psi$,
with some initial displacement $\psi_0$ at $\phi=\phi_c$.
This is justified as long as one of the small patch with $\psi = \psi_0$ expands later and fills 
observable scales of the Universe due to subsequent inflationary expansion.
 
 In the next section, we study the hybrid inflation model in detail in an analytical way,
 and calculate the scalar spectral index and tensor-to-scalar ratio.

 %%%%%%%%%%%%%%%%%%%%%%%%%%%%%%%%%%%%%%%%%%%%%% 
 \section{Dynamics of waterfall field}
 %%%%%%%%%%%%%%%%%%%%%%%%%%%%%%%%%%%%%%%%%%%%%%

We introduce real scalar fields $\phi$ and $\psi$ having the scalar potential of the following form : 
\begin{equation}
        V = \Lambda^4 \left[ \left( 1-\frac{\psi^2}{v^2} \right)^2 + \frac{\phi^2}{\mu^2} + \frac{\phi^2\psi^2}{w^4}   \right].  \label{potential}
\end{equation}
This kind of potential is often realized in supersymmetric (SUSY) theories~\cite{Copeland:1994vg,Dvali:1994ms}.
In SUSY case, $\phi$ and $\psi$ should be regarded as complex scalars and $w=v$ holds.\footnote{
        In SUSY, the potential is actually much more complicated due to the Coleman-Weinberg correction, 
        supergravity correction, SUSY breaking effects
        and so on~\cite{Dvali:1994ms,Linde:1997sj,Senoguz:2004vu,BasteroGil:2006cm,Nakayama:2010xf}. 
}
It is seen that the mass of the waterfall field $\psi$ depends on the value of $\phi$,
and it becomes tachyonic at $\phi=\phi_c$, where
\begin{equation}
        \phi_c = \frac{\sqrt 2  w^2}{v}.
\end{equation}
Therefore, for $\phi>\phi_c$, $\psi$ is stabilized at $\psi=0$ and the potential for $\phi$ is flat and inflation occurs,
and $\psi$ begins to roll down the potential at $\phi=\phi_c$
towards the the potential minimum $\phi=0$ and $\psi = v$.
In order for inflation to occur, $\mu \gg M_P$ is required, since otherwise $\phi$ rolls down too fast.
Also we assume $v < M_P$ and $\phi_c< M_P$ in the following analysis.
Otherwise, chaotic inflation takes place, as briefly discussed in Appendix.

It is often assumed that this waterfall phase transition is sudden and inflation soon ends at $\phi = \phi_c$.
In the following we will study the dynamics around the waterfall point in detail.
Hereafter we define the ``waterfall regime'' as the region where $\phi < \phi_c$.

The equations of motion under the slow-roll approximation are given by
\begin{gather}
	3H\dot\phi = -\frac{2\phi \Lambda^4}{\mu^2}\left( 1+ \frac{2\mu^2\psi^2}{v^2 \phi_c^2} \right),
	\label{dotphi}  \\
	3H\dot\psi = -\frac{4\psi \Lambda^4}{v^2}\left( \frac{\phi^2-\phi_c^2}{\phi_c^2}+ \frac{\psi^2}{v^2} \right).
	\label{dotpsi}
\end{gather}
Slow-roll parameters are directly calculated as
\begin{equation}
        \epsilon_\phi = \frac{1}{2}M_P^2 \left( \frac{V_\phi}{V} \right)^2 =
        \frac{2\phi^2 M_P^2}{\mu^4}\left( 1+\frac{2\mu^2\psi^2}{\phi_c^2 v^2} \right)^2,  \label{eps_phi}
\end{equation}
\begin{equation}
        \epsilon_\psi = \frac{1}{2}M_P^2 \left( \frac{V_\psi}{V} \right)^2 =
         \frac{8\psi^2 M_P^2}{v^4}\left( \frac{\phi^2-\phi_c^2}{\phi_c^2} 
                +\frac{\psi^2}{v^2} \right)^2,        \label{eps_psi}
\end{equation}
\begin{equation}
        \eta_{\phi\phi} = M_P^2 \frac{V_{\phi \phi} }{V} =
        \frac{2M_P^2}{\mu^2}\left(1+ \frac{2\mu^2\psi^2}{\phi_c^2 v^2} \right),     \label{eta_phi}
\end{equation}
\begin{equation}
        \eta_{\phi\psi} = M_P^2 \frac{V_{\phi \psi} }{V} = \frac{8\phi \psi M_P^2}{v^2 \phi_c^2},
\end{equation}
\begin{equation}
        \eta_{\psi\psi} = M_P^2 \frac{V_{\psi \psi} }{V} =
         \frac{4M_P^2}{v^2}\left(  \frac{\phi^2-\phi_c^2}{\phi_c^2} + \frac{3\psi^2}{v^2} \right).   \label{eta_psi}
\end{equation}
Here $V_\phi \equiv \partial V / \partial \phi$ and so on, and $M_P = 2.4\times 10^{18}$GeV is the reduced Planck scale.
The true ends of inflation is at the point where the slow-roll conditions are violated,
and what we are concerning is the dynamics between the critical point $\phi = \phi_c$ and this inflation end point.

\subsection{Classical solution}

For convenience, we parameterize $\phi$ and $\psi$ as
\begin{equation}
        \phi = \phi_c e^{\xi}, ~~~\psi = \psi_0 e^\chi.
\end{equation}
Thus, $\xi=\chi=0$ at the critical point and $\xi < 0$ in the waterfall regime.
We can set $|\xi | \ll 1$ in all the following analyses.
A typical initial displacement $\psi_0$ is of order of the Hubble scale during inflation, $H_{\rm inf} \sim \Lambda^2 / M_P$,
because the $\psi$ field is nearly massless around the critical point and obtains a quantum fluctuation.
Hereafter, we follow the scalar dynamics classically with initial condition of $\psi=\psi_0$.
In the case that sufficiently long inflation occurs during the waterfall regime, this treatment is justified 
because initially spatially small patch with $\psi = \psi_0$ covers the whole observable Universe.
Actually results do not depend much on the value of $\psi_0$.
Otherwise, effects of $\psi$ on the curvature perturbation are more complicated.
But still our analytical results are applied for small region of the Universe.

The scalar field dynamics at the waterfall regime is divided into three stages, which we call phase 0, phase 1 and phase 2
in the chronological order.
The definitions of each stage are as follows.
\begin{itemize}
\item Phase 0 : The second term in Eq.~(\ref{dotpsi}) is dominant.
\item Phase 1 : The first term in Eq.~(\ref{dotpsi}) is dominant.
\item Phase 2 : The second term in Eq.~(\ref{dotphi}) is dominant.
\end{itemize}
At the phase 0, the second term in Eq.~(\ref{dotpsi}) dominates the dynamics of $\psi$ direction
since $\phi \simeq \phi_c$ at the very beginning of the waterfall regime and the first term in Eq.~(\ref{dotpsi}) is neglected.
Then $\psi$ gradually decreases during this phase.
For simplicity, we assume 
\begin{equation}
        \frac{\sqrt{2} \mu \psi_0}{\phi_c v} \ll 1,    \label{assump}
\end{equation}
as an initial condition, and the $\phi$ motion is determined by the first term in Eq.~(\ref{dotphi}) at this stage.
Thus $\phi$ also gradually decreases and begins to deviate from $\phi_c$.
At the phase 1, the first term in Eq.~(\ref{dotpsi}) becomes dominant and $\psi$ increases,
but still the second term in Eq.~(\ref{dotphi}) is small enough to be neglected.
Finally at the phase 2, the second term in Eq.~(\ref{dotphi}) comes to dominate the dynamics of $\phi$.

To be more precise, after the dynamics enters the phase 1, the second term in Eq.~(\ref{dotpsi})
again grows and scalar fields may reach the temporal minimum where 
the r.h.s. of Eq.~(\ref{dotpsi}) is zero. Then the scalar fields track the temporal minimum.
This may happen at the phase 1 or phase 2 during the slow-roll regime, depending on parameters.

Slow-roll conditions are violated somewhere during these processes at $\phi=\phi_{\rm end}$,
and it is not evident whether a large amount of e-folding number is spent for $\phi_{\rm end} < \phi < \phi_c$.
In the following, we closely look into the dynamics of scalar fields at each of these stages
and find the condition for which a sufficient amount of inflation takes place during the waterfall regime.

%Since we are interested in the region $\psi \ll M$, we can always neglect second terms of Eqs.~(\ref{eps_psi}) and (\ref{eta_psi}).
%In the following we divide the waterfall regime into the two stages.
%At the first stage, $\psi$ takes a small value and hence the following approximation holds,
%%
%\begin{equation}
    %    \frac{\sqrt{2} \mu \psi}{\phi_c M} \ll 1,    \label{assump}
%\end{equation}
%%
%so that the second terms in Eqs.~(\ref{eps_phi}) and (\ref{eta_phi}) are neglected.
%At the second stage, $\psi$ takes a large value and the inequality is inverted.

\subsubsection{Phase 0}   \label{sec:phase0}

At this stage of the waterfall regime, slow-roll parameters are simplified as
\begin{equation}
        \epsilon_\phi = \frac{2\phi^2 M_P^2}{\mu^4},
\end{equation}
\begin{equation}
        \epsilon_\psi =  \frac{8\psi^6 M_P^2}{v^8},
\end{equation}
\begin{equation}
        \eta_{\phi\phi} = \frac{2M_P^2}{\mu^2},
\end{equation}
\begin{equation}
        \eta_{\psi\psi} = \frac{12\psi^2 M_P^2}{v^4}.
\end{equation}
Thus we assume $\psi_0 \ll v^{4/3}M_P^{-1/3}$ and $\psi_0 \ll v^{2}M_P^{-1}$
so that slow-roll conditions are satisfied initially.
The trajectory is determined from the slow-roll equation of motion, which can be rewritten as
\begin{equation}
	\frac{d\xi}{d\chi} = \frac{v^4}{2\mu^2 \psi_0^2}e^{-2\chi}.
\end{equation}
Solving this equation, we obtain
\begin{equation}
	\xi = \frac{v^4}{4\mu^2\psi_0^2}(1-e^{-2\chi}).   \label{traj0}
\end{equation}
From the slow-roll equation of motion
\begin{equation}
        3H \dot \xi = -\frac{1}{\phi_c}\frac{\partial V}{\partial \phi},    \label{dotxi}
\end{equation}
we find 
\begin{equation}
        \xi(N) = -\frac{2N M_P^2}{\mu^2}.
\end{equation}
Here $N$ counts the e-folding number after the critical point :
$N=0$ at the critical point and takes a positive value after that.

The phase 0 ends and enters the phase 1 at
\begin{equation}
	\xi = -\frac{\psi_0^2}{2v^2}e^{2\chi}.   \label{tempmin}
\end{equation}
From Eqs.~(\ref{traj0}) and (\ref{tempmin}), we find that
the phase 0 connects to phase 1 at
\begin{gather}
	\xi_1 = \frac{v^4}{8\mu^2\psi_0^2}\left[ 1-\sqrt{1+ \frac{8\mu^2 \psi_0^4}{v^6}}  \right], \\
	\chi_1 = \frac{1}{2}\ln \left(-\frac{2v^2 \xi_1}{\psi_0^2}\right).
\end{gather}
Approximately we have
\begin{equation}
	\xi_1 \simeq 
	\left \{ \begin{array}{ll}
		\displaystyle
		-\frac{v}{2\sqrt 2 \mu} &~~~{\rm for}~~~\frac{8\mu^2 \psi_0^4}{v^6} \gg 1 \\
		\displaystyle
		-\frac{\psi_0^2}{2v^2}+\frac{\mu^2\psi_0^6}{v^8} &~~~{\rm for}~~~\frac{8\mu^2 \psi_0^4}{v^6} \ll 1,
	\end{array}
	\right.
\end{equation}
\begin{equation}
	\chi_1 \simeq 
	\left \{ \begin{array}{ll}
		\displaystyle
		-\frac{1}{2}\ln\left(\frac{v^3}{\sqrt2 \mu\psi_0^2}\right) &~~~{\rm for}~~~\frac{8\mu^2 \psi_0^4}{v^6} \gg 1 \\
		\displaystyle
		-\frac{\mu^2\psi_0^4}{v^6}&~~~{\rm for}~~~\frac{8\mu^2 \psi_0^4}{v^6} \ll 1.
	\end{array}
	\right.
\end{equation}
Therefore, the e-folding number spent for the phase 0 is estimated as
\begin{equation}
	N_0 = -\frac{\mu^2}{2M_P^2}\xi_1\simeq% \leq \frac{1}{4\sqrt 2}\frac{\mu M}{M_P^2}.
	\left \{ \begin{array}{ll}
		\displaystyle
		\frac{\mu v}{4\sqrt 2 M_P^2} &~~~{\rm for}~~~\frac{8\mu^2 \psi_0^4}{v^6} \gg 1 \\
		\displaystyle
		\frac{\mu^2\psi_0^2}{4M_P^2 v^2}&~~~{\rm for}~~~\frac{8\mu^2 \psi_0^4}{v^6} \ll 1.
	\end{array}
	\right.
\end{equation}
Therefore the duration of phase 0 is sufficiently small for $\mu v \ll M_P^2$.

%%%%%%%%%%%%%%%%%%%%
\begin{figure}
\includegraphics[scale=0.5]{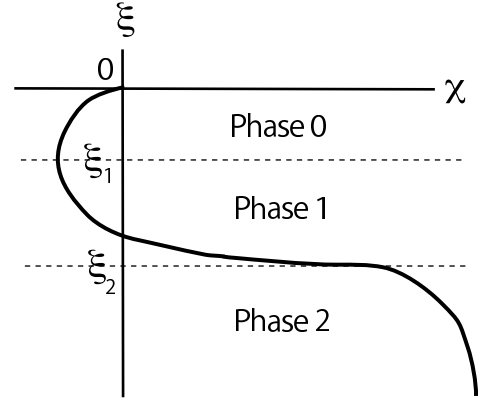}
\caption{
        Schematic picture for the trajectory of the $\xi$ and $\chi$.
 }
\label{fig:schem}
\end{figure}
%%%%%%%%%%%%%%%%%%%%

\subsubsection{Phase 1}    \label{sec:phase1}

The phase 1 is defined as the region where the r.h.s. of Eq.~(\ref{dotpsi}) is dominated by the first term and
\begin{equation}
        \frac{\sqrt{2} \mu \psi}{\phi_c v} \ll 1.    \label{assump1}
\end{equation}
At this stage of the waterfall regime, slow-roll parameters are simplified as
\begin{equation}
        \epsilon_\phi = \frac{2\phi^2 M_P^2}{\mu^4},
\end{equation}
\begin{equation}
        \epsilon_\psi =  \frac{8\psi^2 M_P^2}{v^4}\left( \frac{\phi^2-\phi_c^2}{\phi_c^2} \right)^2 \sim  \frac{32\psi^2 M_P^2 \xi^2}{v^4},
\end{equation}
\begin{equation}
        \eta_{\phi\phi} = \frac{2M_P^2}{\mu^2},
\end{equation}
\begin{equation}
        \eta_{\psi\psi} = \frac{4M_P^2}{v^2}\left(  \frac{\phi^2-\phi_c^2}{\phi_c^2} \right) \sim   \frac{8M_P^2\xi}{v^2}.
\end{equation}
%%
%It is soon found that the slow-roll condition on $\eta_{\psi\psi}$ is violated at
%%
%\begin{equation}
%       \xi = \xi_c = -\frac{M^2}{8M_P^2}.
%\end{equation}
%%
%This is the point where the inflation ends.
We find the trajectory of the scalar fields from the relation
\begin{equation}
        \frac{d\xi}{d\chi} = \frac{v^2}{4\mu^2 \xi},
\end{equation}
which is derived from the equation of motion.  
Matching to the phase 0 solution at $\xi=\xi_1$, we find that $\chi$ is given by
\begin{equation}
        \xi^2 = \xi_1^2 + \frac{v^2}{2\mu^2}(\chi-\chi_1).    \label{traj1}
\end{equation}

The phase 1 connects to phase 2 at $\chi=\chi_2$ defined by
\begin{equation}
        \frac{\sqrt{2} \mu \psi_0 e^{\chi_2}}{\phi_c v} = 1
        \leftrightarrow \chi_2  = \ln  \left(\frac{\phi_c v}{\sqrt 2 \mu\psi_0}\right).    \label{chi2}
\end{equation}
From the assumption (\ref{assump}), we have $\chi_2 > 0$.

Here we must check whether the assumption that the second term in Eq.~(\ref{dotpsi}) is negligible
during the phase 1 $(|\xi_1|< |\xi| < |\xi_2|)$.
If the second term in Eq.~(\ref{dotpsi}) again becomes efficient,
the $\psi$ field is trapped at the temporal minimum of the potential, 
%$\psi_{\rm min}^2 = -2\xi M^2$, 
and then tracks the temporal minimum after that.
The trajectory of the temporal minimum, $\partial V/\partial\psi=0$, is given by
\begin{equation}
	\xi = -\frac{\psi_0^2}{2v^2}e^{2\chi}.   \label{trajmin}
\end{equation}
Let us denote the point where the trajectory (\ref{traj1}) crosses the temporal minimum (\ref{trajmin}) 
by $\xi_{*1}$.
If $|\xi_{*1}| > |\xi_2|$, the fields do not reach the temporal minimum before they enter in the phase 2.
This condition is rewritten as
\begin{equation}
	\chi_2 > \frac{\phi_c^4}{8\mu^2 v^2}.
\end{equation}
If this is satisfied, the trajectory (\ref{traj1}) is valid during the phase 1.
In this case we have $\xi_2 = -cv/\mu$ with $c = \sqrt{\chi_2/2}$ : we call this case as phase 1-(a).
Otherwise, the fields are trapped at the temporal minimum of the potential at 
$\xi=\xi_{*1}$ which satisfies $|\xi_1|<|\xi_{*1}|<|\xi_2|$ : we call this case as phase 1-(b).
%%
%\begin{equation}
%	\xi_* = \frac{\phi_c^2}{2M^2}\left[ -1\pm \sqrt{1+(2\chi_2-1)\frac{\phi_c^4}{\mu^2 M^2}}\right] .
%\end{equation}
%%
In this case the trajectory is given by Eq.~(\ref{trajmin}) for $|\xi|>|\xi_{*1}|$.
and we have $\xi_2 = -\phi_c^2/(4\mu^2)$.

To summarize, the phase 1-(a) trajectory is given by
\begin{gather}
	\xi^2 = \xi_1^2 + \frac{v^2}{2\mu^2}(\chi-\chi_1)~~~{\rm for}~~~|\xi_1|<|\xi|<|\xi_2|,
\end{gather}
where $\xi_2=-cv/\mu$ if $\chi_2 > \phi_c^4/(8\mu^2 v^2)$, and the phase 1-(b) trajectory is given by
\begin{gather}
	\xi^2 = \xi_1^2 + \frac{v^2}{2\mu^2}(\chi-\chi_1)~~~{\rm for}~~~|\xi_1|<|\xi|<|\xi_{*1}|, \\
	\xi = -\frac{\psi_0^2}{2v^2}e^{2\chi} ~~~~{\rm for}~~~|\xi_{*1}|<|\xi|<|\xi_2|,
\end{gather}
where $\xi_2=-\phi_c^2/(4\mu^2)$ if $\chi_2 < \phi_c^4/(8\mu^2 v^2)$.

The e-folding number during the phase 1 is also estimated from the same equation of motion (\ref{dotxi}).
In the phase 1, we obtain
\begin{equation}
	\xi(N) = \xi_1 -\frac{2M_P^2}{\mu^2}(N-N_0).
\end{equation}
Therefore the e-folding number during the phase 1 is given by
\begin{equation}
	\Delta N_1 =N_1-N_0 = \frac{\mu^2}{2M_P^2}(\xi_1 - \xi_2).
\end{equation}
Here $N_1$ is given by
\begin{equation}
	N_1 \simeq 
	\left \{ \begin{array}{ll}
		\displaystyle
		 \frac{\sqrt \chi_2}{2\sqrt 2}\frac{\mu v}{M_P^2}  &~~~{\rm for}~~~\chi_2 > \frac{\phi_c^4}{8\mu^2 v^2} \\
		 \displaystyle
		 \frac{1}{8}\frac{\phi_c^2}{M_P^2} &~~~{\rm for}~~~\chi_2 < \frac{\phi_c^4}{8\mu^2 v^2}.
	\end{array}
	\right.,
\end{equation}
Thus we need at least $\mu v \gg M_P^2$ for long enough inflation to occur at this stage.

One can easily find that at $N \sim \mu v /M_P^2$, $\chi$ becomes close to unity 
if $\chi_2 \sim \mathcal O(1)> \phi_c^4/(8\mu^2 v^2)$.
Then, $\psi$ exponentially grows up, and the approximation (\ref{assump1}) soon breaks down,
hence it goes into the second stage.
At this transition point $N \sim \mu v/M_P^2$, $\eta_{\psi\psi} \sim -8M_P^2 /(\mu v)$.
Thus inflation does not end at the first stage of the waterfall regime as long as $\mu v \gg 8M_P^2$.
In the opposite case $\mu v \ll 8M_P^2$, inflation terminates at $\xi = -M^2/8M_P^2$ during this stage.
But in this case, the e-folding number during the waterfall regime is estimated as
$N = \mu^2 v^2/(16M_P^4) \lesssim 1$.
Thus, in order for sufficient inflation to occur during the waterfall regime,
we need at least $\mu v \gg 8M_P^2$.
In the opposite case $\chi_2 < \phi_c^4/(8\mu^2 v^2)$, the duration is sufficiently short 
as long as $\phi_c \ll M_P$.

%%%%%%%%%%%%%%%%%%%%
\begin{figure}[ht]
\includegraphics[scale=0.5]{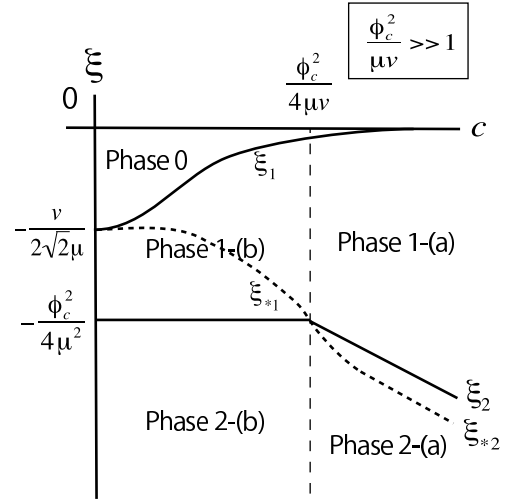}
\caption{
        Schematic presentation of the classification of phase 0-2 for $\phi_c^2/(\mu v) \gg 1$.
        The horizontal axis is $c = \sqrt{\chi_2/2}$ where $\chi_2$ is defined by Eq.~(\ref{chi2}).
        The vertical axis is $\xi$, which starts from 0 and moves down on the figure perpendicularly.
        The dotted line represents $\xi_{*1}$ and $\xi_{*2}$, 
        below which scalars are trapped by the temporal minimum.
 }
\label{fig:phase1}
\end{figure}
%%%%%%%%%%%%%%%%%%%%
%%%%%%%%%%%%%%%%%%%%
\begin{figure}[ht]
\includegraphics[scale=0.5]{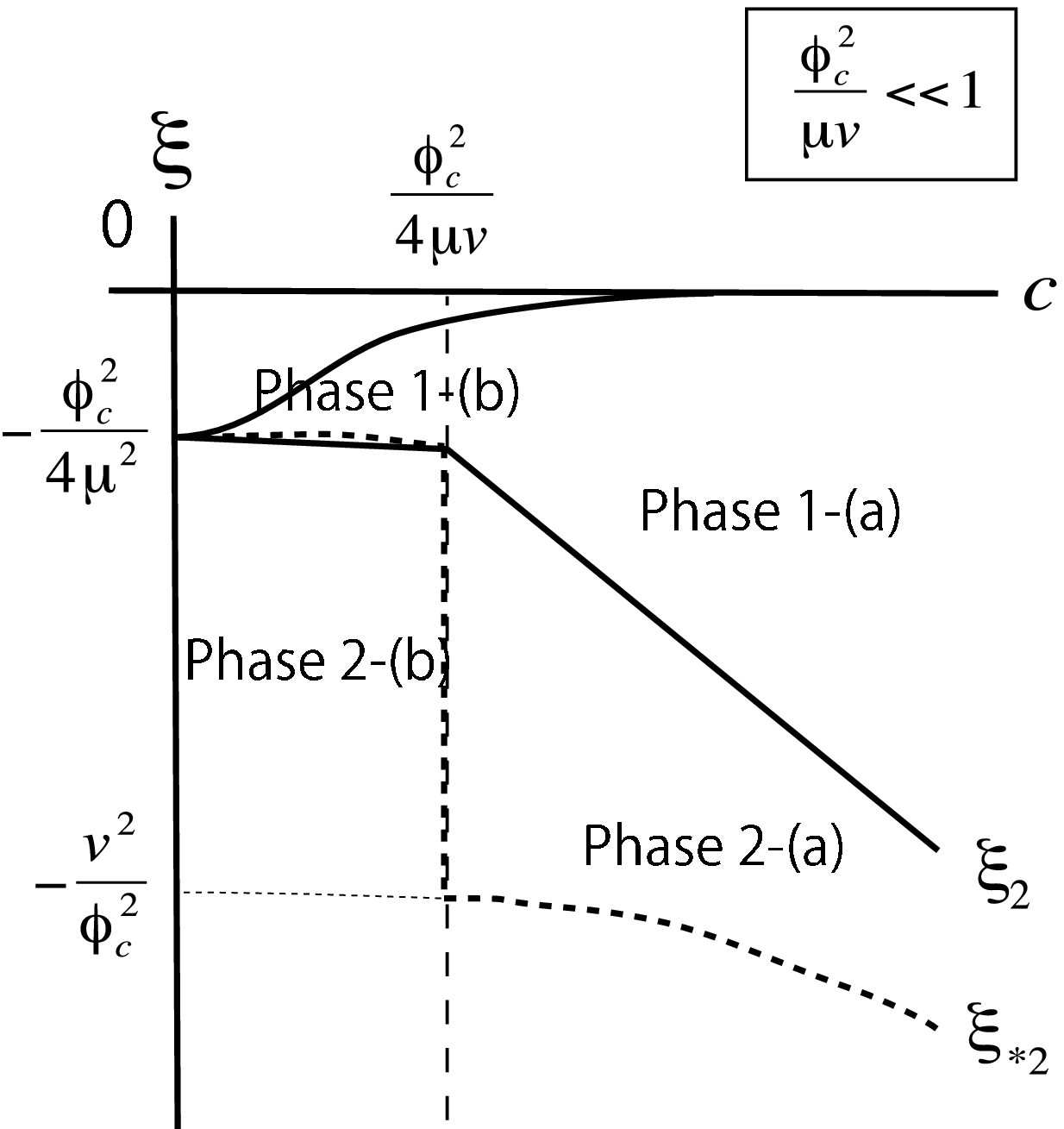}
\caption{
        Same as Fig.~\ref{fig:phase1}, but for $\phi_c^2/(\mu v) \ll 1$.
 }
\label{fig:phase2}
\end{figure}
%%%%%%%%%%%%%%%%%%%%

\subsubsection{Phase 2}  \label{sec:phase2}

The phase 2 is defined as the region where
\begin{equation}
        \frac{\sqrt{2} \mu \psi}{\phi_c v} \gg 1.    \label{assump2}
\end{equation}
At this stage, slow-roll parameters are given by
\begin{equation}
        \epsilon_\phi \simeq \frac{8M_p^2 \psi^4}{\phi_c^2 v^4},
\end{equation}
\begin{equation}
        \epsilon_\psi  \simeq  \frac{32\psi^2 M_P^2 \xi^2}{v^4},
\end{equation}
\begin{equation}
        \eta_{\phi\phi} \simeq \frac{4M_P^2 \psi^2}{\phi_c^2 v^2},
\end{equation}
\begin{equation}
        \eta_{\psi\psi}  \simeq   \frac{8M_P^2\xi}{v^2}.
\end{equation}
%%
%From this we see that the slow-roll condition on $\eta_{\psi\psi}$ is violated at
%%
%\begin{equation}
%        \xi = \xi_c = -\frac{v^2}{8M_P^2}.
%\end{equation}
%%
%This is the point where the inflation ends.
We can also solve the slow-roll equation of motion at this stage.
In this limit, the scalar field trajectory is found from the relation
\begin{equation}
        \frac{d\xi}{d\chi} = \frac{\psi_0^2}{2\phi_c^2}\frac{e^{2\chi}}{\xi}.
\end{equation}
Matching to the phase 1 at $(\xi, \chi) = (\xi_2, \chi_2)$, we obtain
\begin{equation}
        \xi^2 = \xi_2^2 + \frac{v^2}{4\mu^2}\left[ e^{2(\chi-\chi_2)}-1 \right].      \label{traj2}
\end{equation}
As already described, if $\chi_2 < \phi_c^4/(8\mu^2 v^2)$, 
the fields are trapped at the temporal minimum before entering the phase 2,
and the trajectory (\ref{tempmin}) connects to the phase 2 : we call this case phase 2-(b).
For $\chi_2 > \phi_c^4/(8\mu^2 v^2)$, the phase 1-(a) trajectory (\ref{traj1}) connects to the phase 2 :
we call this case phase 2-(a).
Even for the phase 2-(a), there is a possibility that the fields are trapped by the temporal minimum
before inflation ends.
From the trajectory (\ref{traj2}) and the temporal minimum (\ref{tempmin}), we find that
the phase 2 trajectory crosses the temporal minimum at $\xi = \xi_{*2}$, where
\begin{equation}
	\xi_{*2} = -\frac{v^2}{2\phi_c^2}\left[ 1+ \sqrt{ 1+(2\chi_2-1)\frac{\phi_c^4}{\mu^2 v^2} } \right].
\end{equation}
Therefore, approximately we have
\begin{equation}
	\xi_{*2} \simeq 
	\left \{ \begin{array}{ll}
		\displaystyle
		-\frac{v^2}{\phi_c^2} &~~~{\rm for}~\frac{\phi_c^4}{\mu^2 v^2}< \chi_2 < \frac{\mu^2 v^2}{\phi_c^4},\\
		\displaystyle
		-\sqrt{\frac{\chi_2}{2}}\frac{v}{\mu}-\frac{v^2}{2\phi_c^2}  ~(\lesssim \xi_2)  &~~~{\rm otherwise}.
	\end{array}
	\right.
\end{equation}
To summarize, the phase 2-(a) trajectory is given by
\begin{gather}
	\xi^2 = \xi_2^2 + \frac{v^2}{4\mu^2}\left[ e^{2(\chi-\chi_2)}-1 \right]   ~{\rm for}~|\xi_2|<|\xi | <|\xi_{2*}| 
	\label{traj2a1} \\
	\xi = -\frac{\psi_0^2}{2v^2}e^{2\chi} ~~~~{\rm for}~~~|\xi_{2*}|<|\xi |,
	\label{traj2a2}
\end{gather}
where $\xi_2=-cv/\mu$ if $\chi_2 > \phi_c^4/(8\mu^2 v^2)$, and the phase 2-(b) trajectory is given by
\begin{gather}
	\xi = -\frac{\psi_0^2}{2v^2}e^{2\chi} ~~~~{\rm for}~~~|\xi_2|<|\xi |,
	\label{traj2b}
\end{gather}
where $\xi_2=-\phi_c^2/(4\mu^2)$ if $\chi_2 < \phi_c^4/(8\mu^2 v^2)$,
until the slow-roll conditions are violated.
Figs.~\ref{fig:phase1} and \ref{fig:phase2} schematically represent these classifications.
The horizontal axis is $c = \sqrt{\chi_2/2}$ and
the vertical axis is $\xi$, which starts from 0 and moves down on the figure perpendicularly.
The dotted line represents $\xi_*$, under which scalars are trapped by the temporal minimum.
For example, in Fig.~\ref{fig:phase1}, we can see that 
the fields are trapped by the temporal minimum when it traverses the dotted line before reaching the phase 2 regime
for $c \ll \phi_c^2/(\mu v) $. This corresponds to the phase 1-(b).

Using the trajectory (\ref{traj2}), the slow-roll equation of motion becomes
\begin{equation}
        3H\dot\xi = -\frac{8\Lambda^4}{v^2}\left( \xi^2 - \xi_2^2 + \frac{v^2}{4\mu^2} \right).
\end{equation}
If $\chi_2 >1/2$, it can be solved analytically as
\begin{equation}
        %\xi(N) = -\frac{f(N)+16c}{f(N)-16c}\xi_a' ,
        \xi(N) = \frac{-(c'-c)f(N)+c'+c}{(c'-c)f(N)+c'+c}\xi_2' 
\end{equation}
where $c=\sqrt{\chi_2/2}$, $\xi_2' = -c' v/\mu$ with $c'=\sqrt{c^2-1/4}$ and
\begin{equation}
        f(N)=\exp \left( \frac{16c'M_P^2}{\mu v}(N-N_1) \right).
\end{equation}
Note that this expression for $\xi(N)$ diverges $(\xi \to -\infty)$ at $N=N_{\rm div}$, where
\begin{equation}
        N_{\rm div} = N_1 + \frac{\mu v}{16M_P^2 c'}\ln\left( \frac{c+c'}{c-c'} \right).
\end{equation}
Before reaching this point, the slow-roll conditions are violated at $N=N_{\rm end}$,
where $N_{\rm end}$ is defined at the point where $\eta_{\psi\psi} = -1$ and $\xi=\xi_{\rm end}$ with
\begin{equation}
        \xi_{\rm end} = -\frac{v^2}{8M_P^2}.
\end{equation}
Explicitly, it is expressed as
\begin{equation}
        N_{\rm end} = N_1 + \frac{\mu v}{16M_P^2 c'}
                 \ln \left( \frac{\xi_{\rm end}-\xi_2'}{\xi_{\rm end}+\xi_2'}\frac{c+c'}{c-c'} \right).   \label{Nc}
\end{equation}

If the trajectory reaches temporal minimum before inflation ends,
the inflation end point is given by
\begin{equation}
	\xi_{\rm end} = 
	\left \{ \begin{array}{ll}
		\displaystyle
		-\frac{\phi_c^2}{8M_P^2} &~~~{\rm for}~v> \sqrt 2 \phi_c,\\
		\displaystyle
		-\frac{v^2}{16M_P^2} &~~~{\rm for}~v< \sqrt 2 \phi_c.
	\end{array}
	\right.
\end{equation}

To summarize, the $\xi$ evolves as a function of $N$ as
\begin{equation}
\begin{split}
        \xi(N) = -\frac{2M_P^2 N}{\mu^2}&~~~{\rm for ~~}  |\xi| < |\xi_2|, \\
        \xi(N) = \frac{-(c'-c)f(N)+c'+c}{(c'-c)f(N)+c'+c}\xi_2'  &~~~{\rm for ~~} |\xi_2| < |\xi| < |\xi_{\rm end}|.      \label{xiN}
\end{split}
\end{equation}

\begin{figure}
\includegraphics[scale=1.2]{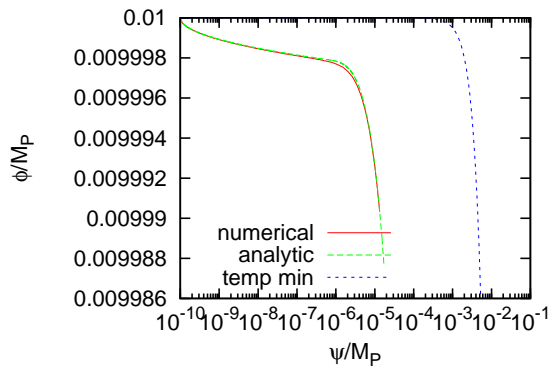}
\caption{
        Trajectory of the $\phi$ and $\psi$ fields.
        We have taken $v=0.1M_P, \phi_c = 0.01M_P, \mu = 10^3 M_P$ and $\psi_0=10^{-10}M_P$.
        This case corresponds to the Phase 1-(a) - Phase 2-(a) solution in Fig.~\ref{fig:phase2}.
        Our analytic solution, plotted until the slow-roll condition is violated, fits very well with the numerical result.
        For comparison, the track of the temporal minimum is also shown.
 }
\label{fig:traj}
\end{figure}

\begin{figure}
\includegraphics[scale=1.2]{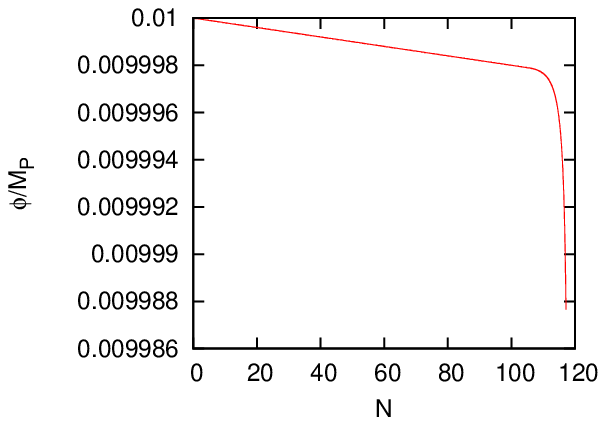}
\caption{
        The $\phi$ as a function of $N$. Parameters are same as those in Fig.~\ref{fig:traj}.
 }
\label{fig:phi}
\end{figure}

\begin{figure}
\includegraphics[scale=1.2]{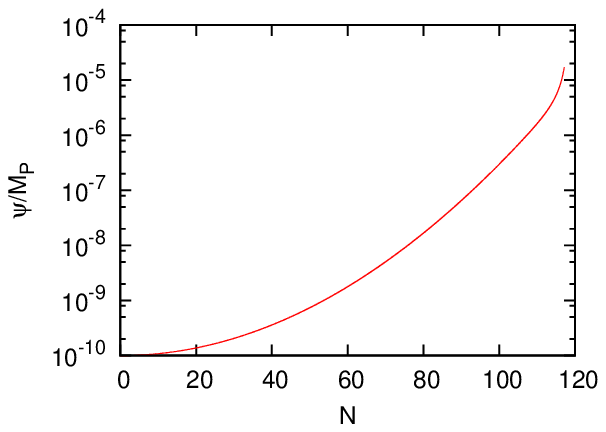}
\caption{
        The $\psi$ as a function of $N$. Parameters are same as those in Fig.~\ref{fig:traj}.
 }
\label{fig:psi}
\end{figure}

A trajectory is  shown in Fig.~\ref{fig:traj}.
We have taken $v=0.1M_P, \phi_c = 0.01M_P, \mu = 10^3 M_P$ and $\psi_0=10^{-10}M_P$.
This case corresponds to the Phase 1-(a) - Phase 2-(a) solution in Fig.~\ref{fig:phase2}.
It is seen that our analytic solution fits very well with the numerical result.
Figs.~\ref{fig:phi} and \ref{fig:psi} show trajectories of the $\phi$ and $\psi$, respectively, as a function of $N$.

\begin{figure}
\includegraphics[scale=1.2]{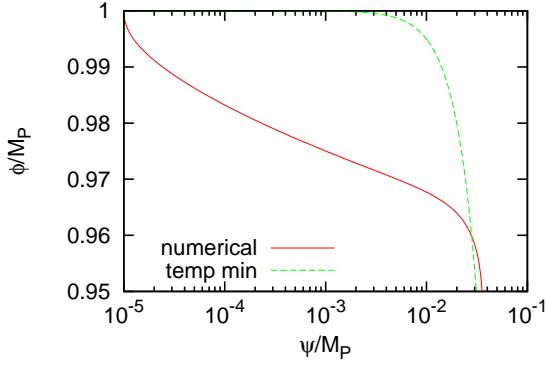}
\caption{
        Trajectory of the $\phi$ and $\psi$ fields.
        We have taken $v=0.1M_P, \phi_c = M_P, \mu = 10 M_P$ and $\psi_0=10^{-5}M_P$.
        This case corresponds to the boundary of Phase 1-(a) - Phase 2-(a) solution and
        Phase 1-(b) - Phase 2-(b) solution in Fig.~\ref{fig:phase1}.
        It is seen that the solution reaches the temporal minimum.
 }
\label{fig:traj2}
\end{figure}

A trajectory for  another set of parameters is shown in Fig.~\ref{fig:traj2}.
We have taken $v=0.1M_P, \phi_c = M_P, \mu = 10 M_P$ and $\psi_0=10^{-5}M_P$.
This case corresponds to the boundary of Phase 1-(a) - Phase 2-(a) solution and
Phase 1-(b) - Phase 2-(b) solution in Fig.~\ref{fig:phase1}.
It is seen that the solution reaches the temporal minimum.

%%%%%%%%%%%%%%%%%%%%
%%%%%%%%%%%%%%%%%%%%

%%%%%%%%%%%%%%%%%%%%
%%%%%%%%%%%%%%%%%%%%

%%%%%%%%%%%%%%%%%%%%
%%%%%%%%%%%%%%%%%%%%

%%%%%%%%%%%%%%%%%%%%
%%%%%%%%%%%%%%%%%%%%

%\subsection{End of inflation}
%Inflation ends at $\eta_{\psi\psi}\simeq -1$.
From Eq.~(\ref{Nc}), we can see that
in order to have enough amount of inflation after waterfall, 
the following condition must be satisfied :
\begin{equation}
        N_{\rm end} \sim N_1 > N_e(\sim 60) \leftrightarrow  \mu M > 2N_e M_P^2.   \label{new}
\end{equation}
This agrees with the result of Clesse~\cite{Clesse:2010iz}.
Thus in this limit, the last $N_e$ e-folds is obtained during the waterfall regime
and actually the waterfall field $\psi$ should be regarded as the inflaton.
It resembles new inflation or type I hilltop model~\cite{Kohri:2007gq}.
In the opposite limit, the waterfall phase transition occurs suddenly 
and the model approaches to the standard hybrid inflation model.

\subsection{Spectral index}

%In the limit (\ref{new}), the $\psi$ field takes a role of the inflaton at least for the observable scale.
Now, we evaluate the spectral index in this model.
First, we must identify the position of the inflaton, $\xi = \xi(N_{\rm end}-N_e)$,
when observable scales left the horizon.
It is given by
\begin{equation}
        \xi(N_{\rm end}-N_e) = \frac{-(c'-c)f(N_{\rm end}-N_e)+c'+c}{(c'-c)f(N_{\rm end}-N_e)+c'+c}\xi_2' ,    \label{xiNe}
\end{equation}
if inflation lasts long enough at $|\xi|>|\xi_2|$, or $N_{\rm end} > N_1 + N_e$.
Otherwise, if $N_{\rm end} < N_1 + N_e$, we obtain
\begin{equation}
        \xi(N_{\rm end}-N_e) = -\frac{2M_P^2(N_{\rm end} - N_e)}{\mu^2}.
\end{equation}

In this model, both $\phi$ and $\psi$ slowly roll down the scalar potential 
and hence the adiabatic field $\sigma$, which is responsible for the curvature perturbation,
should be a combination of these fields.
The scalar spectral index in this case is calculated from~\cite{Gordon:2000hv,Wands:2002bn}
\begin{equation}
        n_S = 1-6\epsilon_\sigma(N_e) + 2\eta_{\sigma\sigma}(N_e),    \label{nS}
\end{equation}
where we have defined slow-roll parameters for the adiabatic field $\sigma$ as
\begin{equation}
        \epsilon_\sigma = \epsilon_\phi + \epsilon_\psi,
\end{equation}
and
\begin{equation}
        \eta_{\sigma\sigma}=\eta_{\phi\phi}\cos^2\theta + 
        2\eta_{\phi\psi}\sin\theta\cos\theta+\eta_{\psi\psi}\sin^2\theta.
\end{equation}
Here, the adiabatic field $\sigma$ is given by
\begin{equation}
        \dot \sigma = \dot \phi \cos\theta  + \dot\psi \sin\theta, 
\end{equation}
where
\begin{equation}
        \cos\theta = \frac{\dot \phi}{\sqrt{ \dot \phi^2 + \dot \psi^2 }},~~~\sin\theta = \frac{\dot \psi}{\sqrt{ \dot \phi^2 + \dot \psi^2 }}.
\end{equation}
Since we already know the analytic solution of the inflaton trajectory,
it is a straightforward task to calculate the scalar spectral index.

\begin{figure}
\includegraphics[scale=0.5]{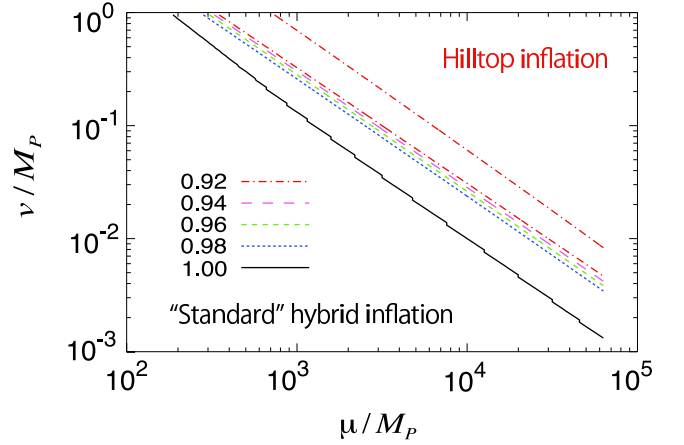}
\caption{
        Contours of the scalar spectral index $n_S$ on $(\mu, v)$-plane. We have taken $\phi_c = 10^{-3}M_P$
        and $\psi_0 = 10^{-15}M_P$.
        The whole region corresponds to the case of Phase 1-(a) - Phase 2-(a) solution in Fig.~\ref{fig:phase2}.
 }
\label{fig:ns}
\end{figure}

Contours of $n_S$ are shown in Fig.~\ref{fig:ns}.
It is seen that in the limit (\ref{new}), the spectral index becomes red,
because the final 60 e-foldings is in the hilltop inflation regime, where the $\psi$ field has a negative curvature.
In the opposite limit, it becomes slightly blue since the $\phi$ field causes 
inflation before the critical point and it has a positive curvature.

A scalar degree of freedom perpendicular to $\sigma$ also has quantum fluctuations
and it is an isocurvature mode~\cite{Kodama:1996jh}.
Whether such an isocurvature mode contributes to the final density perturbation or not
depends on the physics of reheating after inflation.
We do not go into detail of this aspect since it is strongly model dependent.
In the simplest case where both $\phi$ and $\psi$ decay into radiation quickly after inflation,
the isocurvature mode has no physical importance.

%%%%%%%%%%%%%%%%%%%%
%%%%%%%%%%%%%%%%%%%%

\subsubsection{Hilltop inflation limit}

Here we derive the spectral index analytically for the hilltop inflation limit, $\mu v \gg M_P^2 N_e$.
First consider the case where inflation ends at the phase 2-(a) ($c \gg \phi_c^2/(\mu v)$).
From Eq.~(\ref{xiNe}) and using $|\xi_{\rm end}| \gg |\xi_2'|$, we find
\begin{equation}
        \xi(N_{\rm end}-N_e) \simeq -\frac{v^2}{8M_P^2 N_e},
\end{equation}
and
\begin{equation}
        \chi(N_{\rm end}-N_e) \simeq \chi_2 + \ln \left( \frac{\mu v}{4M_P^2 N_e} \right). 
\end{equation}
By substituting them into Eqs.~(\ref{eps_phi})-(\ref{eta_psi}), we obtain
\begin{equation}
        \epsilon_\psi (N_e) \simeq 2 \epsilon_\phi (N_e) \simeq \frac{\phi_c^2 v^4}{64 M_P^6 N_e^4}.
\end{equation}
This means that both $\phi$ and $\psi$ significantly contributes to the adiabatic field.
Other slow-roll parameters are calculated as
\begin{gather}
        \eta_{\phi\phi} (N_e) \simeq \frac{v^2}{8M_P^2 N_e^2}, \\
        \eta_{\phi\psi} (N_e) \simeq \frac{\sqrt 2}{N_e},\\
        \eta_{\psi\psi} (N_e)\simeq -\frac{1}{N_e}.
\end{gather}
Therefore, in this limit, the spectral index (\ref{nS}) is calculated as
\begin{equation}
        n_S \simeq 1 - \frac{4}{N_e}.
\end{equation}
This gives $n_S \simeq 0.92$ for $N_e = 50$ and $n_S \simeq 0.933$ for $N_e = 60$.
From this expression it is clear that the spectral index becomes red in this regime.
This is seen in Fig.~\ref{fig:ns}.

In the case where the last $N_e$ e-foldings occur in the phase 2-(b) regime 
($\mu v \gg M_P^2 N_e$ and $c \ll \phi_c^2/(\mu v)$), we find
\begin{equation}
        \xi(N_{\rm end}-N_e) \simeq -\frac{v^2}{8M_P^2 N_e},
\end{equation}
and
\begin{equation}
        \chi(N_{\rm end}-N_e) \simeq \frac{1}{2}\ln\left( \frac{v^4}{4\psi_0^2 M_P^2 N_e} \right). 
\end{equation}
Using them, we obtain following paramters,
\begin{equation}
	\epsilon_\phi \simeq \frac{v^4}{2\phi_c^2 M_P^2 N_e^2}, ~~~\epsilon_\psi \simeq 0,
\end{equation}
and
\begin{equation}
	\eta_{\phi\phi}\simeq \frac{v^2}{\phi_c^2 N_e}.
\end{equation}
Therefore, the scalar spectral index in this case is given by
\begin{equation}
	n_S \simeq 1 + \frac{2v^2}{\phi_c^2 N_e}.
\end{equation}

\subsubsection{``Standard'' hybrid inflation limit}

On the other hand, if $\mu v \ll M_P^2 N_e$, 
the e-folding number during the waterfall regime is negligibly small and
a conventional picture for the hybrid inflation is recovered.
In this case, we easily find $\phi(N_e) \simeq \phi_c$ and hence
slow-roll parameters are given by
\begin{gather}
        \epsilon_\phi (N_e) \simeq \frac{2\phi_c^2M_P^2}{\mu^4},\\
        \eta_{\phi\phi} (N_e) \simeq \frac{2M_P^2}{\mu^2}.
\end{gather}
Therefore, the scalar spectral index is given by
\begin{equation}
        n_S \simeq 1 + \frac{4M_P^2}{\mu^2} \simeq 1.
\end{equation}
It tends to make the spectral index slightly blue tilted.
These features are clearly found in Fig.~\ref{fig:ns}.

\subsection{WMAP normalization and tensor-to-scalar ratio}

%%%%%%%%%%%%%%%%%%%%
\begin{figure}
\includegraphics[scale=0.5]{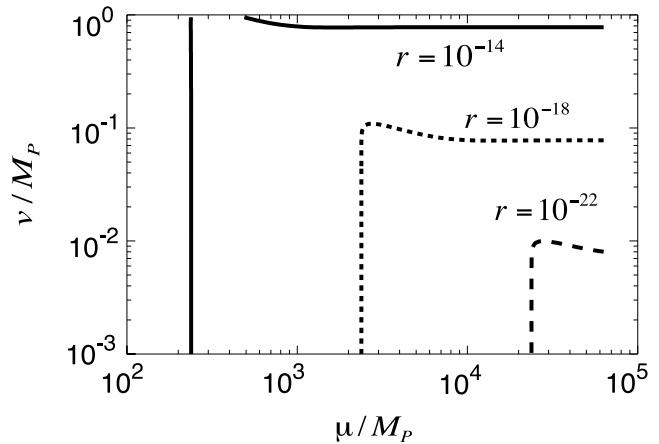}
\caption{
        Contours of the tensor-to-scalar ratio $r$ on $(\mu, v)$-plane. We have taken $\phi_c = 10^{-3}M_P$
        and $\psi_0 = 10^{-15}M_P$.
 }
\label{fig:r}
\end{figure}
%%%%%%%%%%%%%%%%%%%%

So far we have not set the overall inflationary scale, $\Lambda$, in Eq.~(\ref{potential}).
It is determined by the condition that the magnitude of the curvature perturbation 
agrees with the observation.
The WMAP normalization reads~\cite{Komatsu:2010fb}
\begin{equation}
        \mathcal P_{\mathcal R} = \frac{1}{24\pi^2 M_P^4} \frac{V}{\epsilon_\sigma} \simeq 2.4\times 10^{-9},
\end{equation}
where $\mathcal P_{\mathcal R}$ denotes the dimensionless power spectrum of the curvature perturbation
at the pivot scale $k=0.002{\rm Mpc}^{-1}$.
This determines the energy scale of inflation, $\Lambda$.

In the hilltop inflation limit, this leads to
\begin{equation}
        \Lambda \simeq 2.1\times 10^{-4} M_P \left( \frac{v}{M_P} \right)
        \left( \frac{\phi_c}{M_P} \right)^{1/2} \left( \frac{50}{N_e} \right).
\end{equation}
It can also be translated into the tensor-to-scalar ratio, $r$, defined as the ratio between
the power spectrum of the tensor perturbation and $\mathcal P_{\mathcal R}$.
It is related to the slow-roll parameter as $r=16\epsilon_\sigma$.
Thus we have
\begin{equation}
        r \simeq 6.0\times 10^{-8} \left( \frac{\phi_c}{M_P} \right)^2 \left( \frac{v}{M_P} \right) ^4\left( \frac{50}{N_e} \right)^4.
\end{equation}

In the ``standard'' hybrid inflation limit, we obtain
\begin{equation}
        \Lambda \simeq 3.3\times 10^{-2} M_P \left( \frac{\phi_c}{M_P} \right)^{1/2} \left( \frac{M_P}{\mu} \right).
\end{equation}
In terms of the tensor-to-scalar ratio $r$, this gives
\begin{equation}
        r \simeq 3.2\times 10^{-3} \left( \frac{\phi_c}{M_P} \right) ^2 \left( \frac{10M_P}{\mu} \right) ^4.
\end{equation}

Contours of $r$ are shown in Fig.~\ref{fig:r}.
Unfortunately, it is so small that future observations will not have a chance to detect it
for most of the parameter space.

%%%%%%%%%%%%%%%%%%%%%%%%%%%%%%%%%%%%%%%%%%%%%% 
 \section{Conclusions}
%%%%%%%%%%%%%%%%%%%%%%%%%%%%%%%%%%%%%%%%%%%%%%

In this paper, we have reanalyzed the hybrid inflation model, in
particular paying attention to the behavior of the waterfall field
after the critical point.  We have derived analytic formulae
describing the precise motion of both $\phi$ and $\psi$ fields.  In
accordance with the result by Clesse~\cite{Clesse:2010iz}, we found
that sufficiently long inflation takes place during the waterfall
regime for some parameter spaces.  Interestingly, in such a case the
scalar spectral index tends to be red, as opposed to the well known
lore that it becomes blue in the hybrid inflation model, and
consistency with current observations become better.  In this limit,
the $\psi$ causes a inflation similar to new inflation.  New inflation
models often have a problem of initial condition, since initially the
inflaton must be placed near the top of the potential.  In the present
model, it is automatically set around there due to the pre-new
inflation dynamics.  In this sense, the potential of the
hybrid-inflation type (\ref{potential}) may be used for the purpose of
providing an appropriate initial condition of the new inflation.

Some comments are in order.
In general, in the hybrid inflation model topological defects such as domain walls are formed
at the waterfall phase transition.
In order to avoid the problem of domain walls, we need some additional assumptions.
One obvious option is to introduce an additional $Z_2$-breaking term in the scalar potential,
which makes domain walls unstable.
Another option is to extend $\psi$ to be a complex scalar and replace $\psi$ in Eq.~(\ref{potential}) with its absolute value.
In this case it is U(1) symmetry that is spontaneously broken after inflation, 
and correspondingly cosmic strings are formed, which is less harmful than the domain walls.
However, in the case where the last 60 e-foldings takes place during the waterfall regime, we do not need such options,
since topological defects are inflated away and no such objects exist in the observable region of the Universe.
Thus, it has advantages from the viewpoint of not only the spectral index but also the domain wall problem.

%%%%%%%%%%%%%%%%%%%%%%%%%%%%%%%%%%%%%%%%%%%%
\begin{acknowledgments}
%%%%%%%%%%%%%%%%%%%%%%%%%%%%%%%%%%%%%%%%%%%%

We would like to thank Fuminobu Takahashi for useful comments on early stage of this project.
This work is supported by Grant-in-Aid for Scientific research from
the Ministry of Education, Science, Sports, and Culture (MEXT),
No. 21111006. K.K. was partly supported by the Center for the
Promotion of Integrated Sciences (CPIS) of Sokendai.

 %%%%%%%%%%%%%%%%%%%%%%%%%%%%%%%%%%%%%%%%%%%%
\end{acknowledgments}
%%%%%%%%%%%%%%%%%%%%%%%%%%%%%%%%%%%%%%%%%%%%

%%%%%%%%%%%%%%%%%%%
\appendix
%%%%%%%%%%%%%%%%%%%

%%%%%%%%%%%%%%%%%%%%%%%%%%%%%%%%%%%%%%%%%%%%
\section{Exact solution}
%%%%%%%%%%%%%%%%%%%%%%%%%%%%%%%%%%%%%%%%%%%%

\subsection{Phase 0 and Phase 1}

Here we derive analytic solutions for describing both phase 0 and phase 1.
In these phases, we have
\begin{equation}
        \frac{\sqrt{2} \mu \psi}{\phi_c v} \ll 1,
\end{equation}
but we keep the $\psi^2$ term in Eq.~(\ref{dotpsi}).
Then the slow-roll equation of motion leads to
\begin{equation}
	\frac{d\chi}{d\xi} = \frac{2\mu^2}{v^2}\left(e^{2\xi}-1+\frac{\psi_0^2 e^{2\chi}}{v^2} \right).
\end{equation}
Integrating this yields
\begin{equation}
	e^{-2\chi} = e^{-2f(\xi)}-\frac{4\psi_0^2\mu^2}{v^4}\int_0^\xi dx e^{-2f(\xi)+2f(x)},
\end{equation}
under the condition $\chi=0$ at $\xi=0$, where
\begin{equation}
	f(\xi)=\frac{2\mu^2}{v^2}\int_0^\xi dx(e^{2x}-1) = \frac{\mu^2}{v^2}(2|\xi|-1+e^{-2|\xi|}).
\end{equation}
For $|\xi| \ll v/\mu (\ll 1)$, it is approximated as
\begin{equation}
	e^{-2\chi} \simeq 1+\frac{4\mu^2}{v^2}|\xi|\left( \frac{\psi_0^2}{v^2}-|\xi| \right),
\end{equation}
and for $v/\mu \ll |\xi| \ll 1$, we have
\begin{equation}
	e^{-2\chi} \simeq e^{-(2\mu\xi/v)^2} + \frac{\psi_0^2}{2|\xi|v^2}\left( 1-e^{-(2\mu\xi/v)^2} \right).   \label{exact2}
\end{equation}
It is seen that, starting from $\chi=0$ and $\xi = 0$, 
$\chi$ decreases first until $|\xi_1| \simeq \psi_0^2/(2v^2)$ if $\psi_0^2 \ll v^3/\mu$, 
or $|\xi_1| \simeq v/\mu$ if $\psi_0^2 \gg v^3/\mu$, and then increases monotonically.
Comparing them with the results of Sec.~\ref{sec:phase0},
we find that these solutions smoothly connect our phase 0 and phase 1 solutions.
The first term in (\ref{exact2}) corresponds to the phase 1 solution (\ref{traj1}).
Also notice that the second term in (\ref{exact2}) comes to dominate during this phase
for $\chi_2 < \phi_c^4/(8\mu^2 v^2)$ (phase 1-(b)), 
and after that the solution coincides with the temporal minimum (\ref{tempmin}).
Thus our early claim is confirmed that scalars are trapped at the temporal minimum
if $\chi_2 < \phi_c^4/(8\mu^2 v^2)$ by using this exact solution.

\subsection{Phase 2}

Let us seek the solution in the opposite limit,
\begin{equation}
        \frac{\sqrt{2} \mu \psi}{\phi_c M} \gg 1,
\end{equation}
without neglecting $\psi^2$ term in Eq.~(\ref{dotpsi}).
The slow-roll equation of motion leads to
\begin{equation}
	\frac{d\chi}{d\xi}e^{2\chi} = \frac{\phi_c^2}{\psi_0^2}\left(e^{2\xi}-1+\frac{\psi_0^2 e^{2\chi}}{v^2} \right).
\end{equation}
The solution to this equation is written as
\begin{equation}
	e^{2\chi} = \frac{v^2}{\psi_0^2}\frac{1-e^{2k\xi}-k(1-e^{2\xi})}{1-k}+Ce^{2k\xi},
\end{equation}
where $k\equiv \phi_c^2/v^2$ and $C$ is an arbitrary constant determined from the initial condition.
Connecting the phase 1-(a) solution at $\xi_2=-cv/\mu$, we find the phase 2-(a) solution 
\begin{equation}
\begin{split}
	\xi^2 = \xi_2^2 + \frac{v^2}{4\mu^2}\left( e^{2(\chi-\chi_2)}-1 \right) &~~{\rm for~~}|\xi| \ll \frac{v^2}{\phi_c^2} \ll 1, \\
	\xi = -\frac{\psi_0^2}{2v^2}e^{2\chi} &~~{\rm for~~}\frac{v^2}{\phi_c^2} \ll |\xi| \ll 1
\end{split}
\end{equation}
for $ \phi_c^4/(\mu^2 v^2)\ll \chi_2 \ll \mu^2 v^2/\phi_c^4$, which coincides with (\ref{traj2a1}) and (\ref{traj2a2}).

On the other hand, we find a phase 2-(b) solution that connects to phase 1-(b), 
when $\chi_2 < \phi_c^4/(8\mu^2 v^2)$, as
\begin{equation}
	\xi = -\frac{\psi_0^2}{2v^2}e^{2\chi},
\end{equation}
and hence tracks the temporal minimum, as was already shown in (\ref{traj2b}).

%%%%%%%%%%%%%%%%%%%%%%%%%%%%%%%%%%%%%%%%%%%%
\section{Chaotic inflation limit}
%%%%%%%%%%%%%%%%%%%%%%%%%%%%%%%%%%%%%%%%%%%%

So far we have focused on the parameter ranges $\phi_c \ll M_P$ and $v \ll M_P$.
In the opposite case $\phi_c \gg M_P$ and/or $v \gg M_P$,
the scalar potential around the minimum $(\phi,\psi)=(0,v)$ is obviously flat beyond the Planck scale.
Thus the final 60 e-folding during the waterfall regime rather looks like chaotic inflation.

We can expand the scalar potential around the minimum by using $\delta \psi \equiv \psi - v$, as
\begin{equation}
	V = \frac{1}{2}m_\phi^2 \phi^2 + \frac{1}{2}m_\psi^2 \delta \psi^2,
\end{equation}
for $|\delta \psi| \ll v$ and $|\phi| \ll \phi_c$, where
\begin{gather}
	m_\phi^2 = \frac{2\Lambda^4}{\mu^2} + \frac{4\Lambda^4}{\phi_c^2},\\
	m_\psi^2 = \frac{8\Lambda^4}{v^2}.
\end{gather}
Therefore, if either $\phi_c \gg M_P$ or $v \gg M_P$ are satisfied, 
the scalar potential is flat beyond $\phi \sim M_P$ or $|\delta\psi| \sim M_P$,
and chaotic inflation along the corresponding direction can take place.
In this case, the scalar spectral index and tensor-to-scalar ratio are estimated to be $n_S \sim 0.96$ and $r\sim 0.16$,
and the WMAP normalization constrains the mass parameter as $m_\phi (m_\psi) \sim 10^{13}$GeV.

%%%%%%%%%%%%%%%%%%%%%%%%%%%%%%%%%%%%%%%%%%%%
   
%%%%%%%%%%%%%%%%%%%%%%%%%%%%%%%%%%%%%%%%%%%%%
 
 \end{document}